\documentclass[prd,onecolumn,showpacs,floatfix,superscriptaddress,nofootinbib]{revtex4-2}
\usepackage{graphicx}
\usepackage{epsfig}
\usepackage{bm}
\usepackage{amssymb}
\usepackage{float}
\usepackage{amsmath}
\usepackage{dcolumn}
\usepackage{cancel}

\usepackage{mathrsfs}
\usepackage{dcolumn}
\usepackage{graphicx}
\usepackage{amsmath}
\usepackage{amsfonts}
\usepackage{amssymb}
\usepackage{microtype}
\usepackage{subfigure}
\usepackage{makeidx}
\usepackage{bm}
\usepackage{epsf}
\usepackage{color}
\usepackage{multirow,dcolumn}
\usepackage{graphicx}
\usepackage{mathrsfs}
\graphicspath{{Images/}}

\def\doi{http://doi.org}

\def\be{\begin{equation*}}
\def\ee{\end{equation*}}

\def\Ref{\ref}

\begin{document}

\title{ $(2+1)$-Dimensional Black Holes in $f(R,\phi)$ Gravity}

\author{Thanasis Karakasis}
\email{thanasiskarakasis@mail.ntua.gr} 
\affiliation{Physics Division,
National Technical University of Athens, 15780 Zografou Campus,
Athens, Greece.}

\author{Eleftherios Papantonopoulos}
\email{lpapa@central.ntua.gr} 
\affiliation{Physics Division,
National Technical University of Athens, 15780 Zografou Campus,
Athens, Greece.}

\author{Zi-Yu Tang}
\email{tangziyu@ucas.ac.cn}
\affiliation{School of Fundamental Physics and Mathematical Sciences, Hangzhou Institute for Advanced Study, UCAS, Hangzhou 310024, China}
\affiliation{School of Physical Sciences, University of Chinese Academy of Sciences, Beijing 100049, China}

\author{Bin Wang}
\email{wang\_b@sjtu.edu.cn}
\affiliation{School of Aeronautics and Astronautics, Shanghai Jiao Tong University, Shanghai 200240, China}
\affiliation{Center for Gravitation and Cosmology, College of Physical Science
and Technology, Yangzhou University, Yangzhou 225009, China}

\vspace{3.5cm}
\begin{abstract}
We consider a $f(R)$ gravity theory in $(2+1)$-dimensions with a self-interacting scalar field non-minimally coupled to gravity. Without specifying the form of the $f(R)$ function, solving the field equations we find that the Ricci scalar receives a non-linear correction term which breaks the conformal invariance and leads to a massless black hole solution. When the non-linear term decouples, we get a well known hairy black hole solution with the scalar field conformally coupled to gravity. We also find that  the entropy of our black hole may be higher than the corresponding conformal  black hole  which indicates that our solution may be thermodynamically preferred.
\end{abstract}

\vspace{3.5cm}
\maketitle

\flushbottom

\tableofcontents

\section{Introduction}

In $(2+1)$-dimensions one of the first exact black hole solutions of General Relativity (GR) is the Ba\~{n}ados,  Teitelboim, and Zanelli (BTZ) black hole \cite{Banados:1992wn, Banados:1992gq}. It is known that in $(2+1)$-dimensions, the Weyl tensor vanishes by definition and in the absence of matter we have $R_{\mu\nu}=0$, therefore we cannot have a black hole solution. To evade this problem, a negative cosmological constant was introduced in the theory allowing black hole solutions to exist in a spacetime with negative constant curvature. Introducing a linear Maxwell field, a black hole solution can also be obtained with a dynamical Ricci scalar, while a conformally invariant non-linear electromagnetic field $\left(F_{\mu\nu}F^{\mu\nu}\right)^{3/4}$ gives a black hole solution with constant Ricci scalar \cite{Cataldo:2000we}.

After the discovery of the BTZ black hole, several $(2+1)$-dimensional black hole solutions have been obtained coupled to a scalar field. In \cite{Martinez:1996gn, Henneaux:2002wm} black holes with an everywhere regular scalar field have been investigated. Then more $(2+1)$-dimensional hairy black hole solutions were discussed  \cite{CMT1,CMT2,Correa:2012rc,Baake:2020tgk,Bravo-Gaete:2014haa,Natsuume,Aparicio:2012yq,Xu:2014uha}.  In \cite{Cardenas:2014kaa} asymptotically AdS $(2+1)$-dimensional black hole solutions with conformally coupled scalar and Abelian gauge fields were found.  In \cite{Xu:2014uka} and \cite{Xu:2013nia} $(2+1)$-dimensional charged black holes with scalar hair were derived, where the scalar potential is not fixed ad hoc but instead derived from Einstein's equations. In \cite{Tang:2019jkn} exact $(2+1)$-dimensional black holes with a non-minimally coupled scalar field were discussed, where the arbitrary coupling constant breaks the conformal invariance. In \cite{Chan:1994qa} and \cite{Chan:1996rd}, static black holes in $(2+1)$-dimensional dilaton gravity and modifications of the BTZ black hole by a dilaton/scalar were investigated. Recently, regular black hole solutions with a real scalar field coupled to the Maxwell field via a duality transformation were constructed \cite{Bueno:2021krl} and $(2+1)$-dimensional regular black holes with nonlinear electrodynamics sources were studied in \cite{Cataldo:2000ns, HabibMazharimousavi:2011gh, He:2017ujy}.

In four dimensions one of the first hairy black holes was derived by Bocharova, Bronnikov and  Melnikov and independently by Bekenstein, called BBMB black hole \cite{BBMB}. The BBMB action consists of GR and a conformally coupled scalar field. The metric has the form of an extremal Reissner-N\"ordstrom spacetime, however the scalar field is divergent at the horizon and it is found to be unstable under scalar perturbations \cite{bronnikov}. A cosmological constant can make the scalar field regular at horizon \cite{Martinez:2002ru} but the solution is still unstable \cite{Harper:2003wt}. Later the MTZ black hole was considered with a scalar potential \cite{Martinez:2004nb}, where the scalar field is finite at horizon and the spacetime is hyperbolic. Interestingly, a charged black hole solution with a scalar field minimally coupled to gravity was found \cite{Martinez:2006an}, in which the mass of the black hole is zero due to the cancellation of the  contributions of the scalar field and the gravitational field. In this solution  the self-interacting potential of the scalar field breaks the conformal invariance of the MTZ black hole. A new class of exact hairy black hole solutions was discussed in \cite{Kolyvaris:2009pc, Kolyvaris:2011fk}. In \cite{Gonzalez:2013aca,Gonzalez:2014tga} neutral and charged black holes with a minimally coupled scalar field have also been found. In \cite{Rinaldi:2012vy, Anabalon:2013oea, Minamitsuji:2013ura} black holes with non-minimal derivative coupling were studied. Black holes in bi-scalar extensions of Horndeski theories were found in \cite{Charmousis:2014zaa}, hairy black hole solutions with time dependent scalar field were discussed in \cite{Babichev:2013cya} and in \cite{Erices:2021uyu} hairy black holes in disformal scalar-tensor gravity theories were discussed. The stability of black holes with non-minimally coupled scalar hair was studied in \cite{Chatzifotis:2021pak}.

The $f(R)$ theories  were mainly introduced in cosmology in an attempt to describe the early and late cosmological history of our Universe  \cite{DeFelice:2010aj,Starobinsky:1980te}.  The consistency with GR   imposed constraints on the choices of the $f(R)$ models \cite{Capozziello:2014zda}. In these theories black hole solutions were found, that deviate from the GR black holes or they possess new properties that are distinguishable from the known GR solutions. Static and spherically symmetric black hole solutions were derived in $(3+1)$-dimensions and $(2+1)$-dimensions \cite{Sebastiani:2010kv,Multamaki:2006zb,Amirabi:2015aya,Hendi:2014wsa,Hendi:2014mba}, while in \cite{Multamaki:2006ym,Nashed:2020kdb,Nashed:2020tbp,Nashed:2019uyi,Nashed:2019tuk,Hurtado:2020gic,Cembranos:2011sr,Jaryal:2021lsu,Nashed:2021jvn,Tang:2019qiy} charged and rotating solutions were found and thin-shells surrounding $f(R)$ black holes with dynamical scalar curvature were discussed \cite{Eiroa:2020xee}. Static and spherically symmetric black hole solutions were investigated  with constant curvature,  with and without electric charge and cosmological constant in \cite{delaCruzDombriz:2009et,Hendi:2011eg,Eiroa:2020dip}. Scalar fields have been recently introduced as matter fields in the context of $f(R)$, exact black hole solutions have been found and the corresponding physical properties have been investigated in \cite{Tang:2020sjs,Karakasis:2021rpn,Karakasis:2021lnq}, while the stability of $f(R)$ black holes was also considered in \cite{Aragon:2020xtm}. The possibility of dressing the black hole with a gravitational hair has also been discussed in \cite{Canate:2017bao, Canate:2015dda}.

In this work, we  generalize the $(2+1)$-dimensional  conformal  black hole solution \cite{Martinez:1996gn} in $f(R)$ gravity. Although our study shows that the conformal invariance has to be broken due to the nonlinear term of the $f(R)$ model, an exact black hole solution can be solved analytically, while the $f(R)$ theory can be obtained asymptotically. When the nonlinear term of the $f(R)$ model decouples, the solution reduces to the conformally dressed black hole with $f(R)=R+2l^{-2}$ (the abbreviation "GR black hole/solution/case" will always refer to the conformally dressed $(2+1)$-dimensional black hole \cite{Martinez:1996gn}) . Besides, the entropy of the $f(R)$ black hole may be higher than the corresponding conformal black hole, which indicates our solution may be thermodynamically preferred.

The paper is organized as follows. In Section \ref{GR} we review the known conformally dressed black hole solution \cite{Martinez:1996gn} for comparison. In Section \ref{sect2} we generalize the  conformally dressed black hole in $f(R)$ gravity. An exact black hole solution is obtained with the $f(R)$ function being solved from the equations asymptotically, while the conformal invariance is broken due to the nonlinear term of the $f(R)$ model. We show that the $f(R)$ model is free of ghosts and tachyonic instabilities. In Section \ref{sect3} we study the thermodynamics of the solution and find that the solution is thermodynamically preferred for most cases. Finally, in Section \ref{sect4} we conclude.

\section{Conformal $(2+1)$-dimensional black hole}
\label{GR}
We begin with a brief review of the conformally dressed black hole derived in \cite{Martinez:1996gn}. The action of the theory consists of the Ricci scalar, a negative cosmological constant, and a conformally coupled scalar field, namely
\begin{equation} S=\frac{1}{2}\int d^3x \sqrt{-g}\Big\{\frac{R +2 l^{-2}}{\kappa} -\partial_{\mu}\phi\partial^{\mu}\phi-\frac{1}{8}R\phi^2 \Big\}~, \end{equation}
where we will use $\kappa = 8\pi G=1$ for simplicity throughout the paper. By variation one can obtain the Einstein equation and the Klein-Gordon equation
\begin{eqnarray}
&&G_{\mu\nu} -g_{\mu\nu}l^{-2} = T_{\mu\nu}~,\label{Einstein1}\\
&&\Box \phi - \cfrac{1}{8}R\phi  =0~, \label{KG1}
\end{eqnarray}
where $G_{\mu\nu}\equiv R_{\mu\nu}-\frac{1}{2} g_{\mu\nu}R$ and the energy-momentum tensor is given by
\begin{equation} T_{\mu\nu} = \partial_{\mu}\phi\partial_{\nu}\phi - \frac{1}{2}g_{\mu\nu}\partial^{\alpha}\phi\partial_{\alpha}\phi + \frac{1}{8}\Big(g_{\mu\nu}\Box - \nabla_{\mu}\nabla_{\nu} + G_{\mu\nu}\Big)\phi^2 ~. \end{equation}
One can prove that by virtue of equation (\Ref{KG1}) the energy-momentum tensor is traceless so that we have a constant Ricci scalar
\begin{equation} R = -\frac{6}{l^2}~.\label{R}\end{equation}
We assume the metric ansatz with $g_{tt}g_{rr}=-1$
\begin{equation}ds^2 = -b(r)dt^2 + b(r)^{-1}dr^2 + r^2 d\theta^2~, \label{metric} \end{equation}
where $b(r)$ is the only degree of freedom and it can be obtained from the Ricci scalar (\ref{R})
\begin{equation} b(r) = \frac{r^2}{l^2}-\frac{c_1}{r}+c_2~, \end{equation}
while from the $tt$ and $rr$ components of the Einstein equation we can get the scalar field
\begin{equation} \phi(r) = \frac{1}{\sqrt{c_3r+c_4}}~.\end{equation}
Substituting them into the $\theta\theta$ component of the Einstein equation and together with the Klein-Gordon equation (\Ref{KG1}) we have
\begin{eqnarray}
&&b(r) = \frac{r^2}{l^2} + \frac{B^2 (-2 B-3 r)}{l^2 r}~,\\
&&\phi(r) = \sqrt{\frac{8B}{r+B}} \label{scalarfield}~,
\end{eqnarray}
where $B=c_4/c_3>0$.

 This solution is regular for any positive $r$, except for a singularity at the origin $r=0$, as can be seen from the divergence of the Kretschmann scalar $K\equiv R_{\mu\nu\sigma\rho}R^{\mu\nu\sigma\rho}$.
The metric function $b(r)$ has only one root which gives the radius of the event horizon $r_h=2B$. The scalar field does not diverge at the event horizon $r_h$ like the BBMB black hole \cite{BBMB} because of the presence of a negative cosmological constant.

The Hawking temperature \cite{Hawking:1975vcx} is given by the Euclidean trick ($t \to -i \tau$)
\begin{equation}
    T_H=\frac{b'(r_h)}{4\pi }=\frac{9 B}{8 \pi  l^2}~. \label{GRTemp}
\end{equation}
Using Wald's formula \cite{Wald:1993nt}, the entropy at the event horizon can also be obtained as
\begin{equation} S = \frac{\mathcal{A}}{4}\left(1-\frac{1}{8}\phi(r_h)^2\right) = \frac{\pi r_h}{3} =\frac{2 \pi  B}{3}~,\end{equation}
where $\mathcal{A}=2\pi r_h$ is the horizon area.

The entropy acquires another term, besides the GR one, that depends on the scalar field, which comes from the non-minimal coupling between matter and curvature. As a result, the entropy is smaller than the corresponding BTZ black hole entropy \cite{Banados:1992wn} which is $\pi r_h/2$, but is positive and finite, while the entropy of the BBMB black hole is infinite due to the divergence of the scalar field at the event horizon.
 The conserved black hole mass can be obtained by using the first law of thermodynamics
\begin{equation} dM = TdS \rightarrow M= \int T(r_h)S'(r_h)dr_h = \frac{3 r_h^2}{32 l^2} = \frac{3 B^2}{8 l^2}~. \label{GRmass}\end{equation}
All thermodynamic quantities grow with the increase of the scalar charge $B$ (or $r_h$), in agreement with those obtained from the Hamiltonian formalism \cite{Martinez:1996gn}.

The scalar field dresses the black hole with a secondary scalar hair, since its charge $B$ is not an independent conserved quantity, as it is related to the conserved mass of the black hole.

\section{$f(R)$ Gravity Black Hole Solution}
\label{sect2}

 In this Section we  extend the conformal black hole solution \cite{Martinez:1996gn} described in the previous Section in $f(R)$ gravity by replacing the Einstein-Hilbert term $R$ with the $f(R)$ function and endowing the scalar field with a self-interacting potential $V(\phi)$ in the action

\begin{equation} S = \frac{1}{2}\int d^3 x \sqrt{-g}\Big\{f(R) - \partial_{\mu}\phi\partial^{\mu}\phi - \cfrac{1}{8}R\phi^2 -2V(\phi)\Big\}~, \label{action}\end{equation}
The field equations that arise from this action are
\begin{eqnarray}
I_{\mu\nu}\equiv && f_R R_{\mu\nu} - \cfrac{1}{2}g_{\mu\nu}f(R) + g_{\mu\nu}\Box f_R -\nabla_{\mu}\nabla_{\nu}f_R = T_{\mu\nu}~, \label{Einstein}\\
&&\Box \phi - \cfrac{1}{8}R\phi - V'(\phi) =0~, \label{KG}
\end{eqnarray}
where $f_R\equiv\cfrac{df(R)}{dR}$ and the energy momentum tensor becomes
\begin{equation} T_{\mu\nu} = \partial_{\mu}\phi\partial_{\nu}\phi - \frac{1}{2}g_{\mu\nu}\partial^{\alpha}\phi\partial_{\alpha}\phi + \frac{1}{8}\Big(g_{\mu\nu}\Box - \nabla_{\mu}\nabla_{\nu} + G_{\mu\nu}\Big)\phi^2 - g_{\mu\nu}V(\phi)~. \end{equation}
The trace of Einstein equation (\Ref{Einstein}) gives
\begin{equation}I^{\mu}_{\,\mu}\equiv 2f_R R - 3 f(R) +4\Box f_R = \phi \Box \phi -R\phi^2/8-6V(\phi)~. \label{T}\end{equation}
Assuming the same metric ansatz (\Ref{metric}), the $tt,rr,\theta\theta$ components of the Einstein equation and the Klein-Gordon equation take the form
\begin{eqnarray}
&&tt: 2 r \left(b' \left(\phi  \phi '-4 f'_R\right)+4 f_R b''-2 b \left(4 f''_R+\phi '^2-\phi \phi ''\right)+4 f-8 V\right)+b' \left(8 f_R+\phi ^2\right)-16 b f'_R+4 b \phi  \phi '=0~, \label{tt1}\\
&&rr: 2 r \left(b' \left(\phi  \phi '-4 f'_R\right)+4 f_R b''+4 b \phi '^2+4 f-8 V\right)+b' \left(8 f_R+\phi ^2\right)-16 b f'_R+4 b \phi \phi '=0~,\label{rr1}\\
&&\theta\theta: r \left(4 b' \left(\phi  \phi '-4 f'_R\right)+\phi ^2 b''-4 b \left(4 f''_R+\phi '^2-\phi  \phi ''\right)+8 f-16 V\right)+16 f_R b'=0~,\label{uu1} \\
&&\text{KG: } b'(r) \phi '(r)+\frac{\phi (r) \left(2 b'(r)+r b''(r)\right)}{8 r}+\frac{b(r) \phi '(r)}{r}+b(r) \phi ''(r)-\frac{V'(r)}{\phi '(r)}=0~, \label{Klein1}
\end{eqnarray}
also the trace  (\Ref{T}) becomes
\begin{equation}I^{\mu}_{\,\mu} \equiv -32 r b' f'_R+32 f_R b'+16 r f_R b''+8 r \phi  b' \phi '+2 \phi ^2 b'+r \phi ^2 b''-32 b f'_R-32 r b f''_R+8 b \phi \phi '+8 r b \phi  \phi ''+24 r f-48 r V=0~. \label{trace1}\end{equation}

 The Klein-Gordon equation can be obtained by taking the covariant derivative of Einstein's equation \cite{Karakasis:2021rpn}. Therefore, we have a system of three independent equations with four unknown functions: the $f(R)$ function, the potential $V(\phi)$, the scalar field $\phi(r)$ and the metric function $b(r)$. We will leave the potential undetermined and solve it from the field equations. We will then check the trace of the energy-momentum tensor. A vanishing trace will indicate that the matter field is conformally coupled to gravity and a scale (if any) is counterbalanced in the action.
From equations (\Ref{tt1}) and (\Ref{rr1}) we can obtain the relation between the gravitational function $f_R(r)$ and the scalar field $\phi(r)$
\begin{equation} 4 f''_R(r)+3 \phi '(r)^2-\phi (r) \phi ''(r) =0~. \label{ttrr}\end{equation}
We can immediately integrate this equation for $f_R(r)$
\begin{equation} f_R(r) = s + \alpha r + \int \int \frac{1}{4}\left(\phi (r) \phi ''(r)-3 \phi '(r)^2 \right) dr dr~, \label{basic}\end{equation}
where $s$ and $\alpha$ are constants of integration. The constant $s$ is the coefficient of the Einstein-Hilbert term, $\alpha$ is related to geometric corrections to Einstein gravity that are encoded in $f(R)$ theories and the last term is generated from the scalar field. It shows that the scalar field gives an immediate modification to the $f(R)$ model if the integrand does not equal zero. To simplify the equations we consider the integrand to be vanishing, i.e. $f_R''(r)=0$, which gives the profile of the scalar field as $\phi(r) = \sqrt{A/(r+B)}$ and $f_R(r) = s+ \alpha r$. Also, in order to make it comparable with the GR case \cite{Martinez:1996gn} we {use $A=8B$ and $s=1$, then the scalar field becomes same with (\Ref{scalarfield}) and
\begin{equation}
f_R(r) = 1 + \alpha r~. \label{f_R1}
\end{equation}
We can immediately integrate $f_R(r)$ with respect to Ricci scalar to obtain the general form of the $f(R)$ theory
\begin{equation} f_R(r) = 1 + \alpha r \to f(R) = R + \alpha \int^{R}r(R) dR +C~, \end{equation}
where $C$ is an integration constant with the unit $[L]^{-2}$, related to the cosmological constant. This expression shows that a geometric correction term appears in addition to the Einstein-Hilbert term, while the scalar field does not appear immediately in the $f(R)$ model as happens in \cite{Karakasis:2021lnq}.

Then we can solve the metric function as
\begin{equation} b(r) = -\frac{3 B^2}{l^2 (\alpha  B+1)^2}-\frac{2 B^3}{l^2 r (\alpha  B+1)}+\frac{6 \alpha  B^2 r}{l^2 (\alpha  B+1)^3}+r^2\left(\frac{1}{l^2} +\frac{6 \alpha ^2 B^2 }{l^2 (\alpha  B+1)^4}\ln \left(\frac{r}{\alpha  l (B+r)+l}\right)\right)~, \label{b} \end{equation}
where $l$ is the AdS radius that appears as an integration constant. We can see that the metric function is well behaved for any $r>0$ if we constrain the parameters $B, \alpha$ to be positive. For this reason we will impose that $\alpha, B>0$.
At large distances, the metric function asymptotes to
\begin{equation} b(r\to \infty) \sim -\Lambda_{\text{eff}} r^2 -\frac{2 B^2}{\alpha  l^2 r} + \mathcal{O}(r^{-2})~,\end{equation}
where the effective cosmological constant that the $f(R)$ theory and the non-minimal coupling generate is given by
\begin{equation} \Lambda_{\text{eff}} = - \left(\frac{1}{l^2}-\frac{6 \alpha ^2 B^2 \ln (\alpha  l)}{l^2 (\alpha  B+1)^4}\right)~. \end{equation}
For vanishing scalar charge $B$, we obtain pure AdS spacetime and we will also consider that $1-6 \alpha ^2 B^2 \ln (\alpha  l)/(\alpha  B+1)^4>0$ in order to have an asymptotically AdS spacetime, so we can compare our solution with \cite{Martinez:1996gn}.
Now, we can obtain the potential from the Klein-Gordon equation
\begin{multline}
V(r)= \frac{B^3}{l^2 (\alpha  B+1)^4} \Bigg( \frac{6 \alpha ^2 \left(B^2+\alpha  (B+r)^3\right) }{(B+r)^3}\ln \left(\frac{r}{\alpha  l (B+r)+l}\right)+\frac{3 \left(\alpha ^2-\alpha ^4 B^2\right)}{B+r}+\frac{3 \alpha ^3 (\alpha  B+1)^2}{\alpha  (B+r)+1}\\+\frac{6 \alpha ^2 B (\alpha  B+1)}{(B+r)^2}+\frac{\alpha  B (\alpha  B+1) (\alpha  B (\alpha  B+5)-2)}{(B+r)^3}+6 \alpha ^3 \ln \alpha l \Bigg)~, \end{multline}
which vanishes at spatial infinity  and as a function of $\phi$ reads
\begin{multline}
V(\phi) = \frac{\alpha  B}{512 l^2 (\alpha  B+1)^4 \left(8 \alpha  B+\phi ^2\right)}\Bigg(\phi ^2 (\alpha  B+1) (3072 \alpha ^2 B^2+\phi ^6 (\alpha  B (\alpha  B+5)-2)+8 \alpha  B \phi ^4 (\alpha  B+1) (\alpha  B+4)\\+192 \alpha
   B \phi ^2 (\alpha  B+1))+6 \alpha  B \left(8 \alpha  B+\phi ^2\right) \left(\left(512 \alpha  B+\phi ^6\right) \ln \left(\frac{B
   \left(-\phi ^2+8\right)}{l \left(8 \alpha  B+\phi ^2\right)}\right)+512 \alpha  B \ln \left(\alpha  l\right)\right)\Bigg)~.
\end{multline}
The Ricci scalar can be obtained from the metric function
\begin{equation}
R(r) =-\frac{6}{l^2} -\frac{6 \alpha  B^2 \left(\alpha  \left(2 \alpha  B^2+9 \alpha  B r+4 B+6 \alpha  r^2+9 r\right)+2\right)}{l^2 r (\alpha  B+1)^3 (\alpha  (B+r)+1)^2}-\frac{36 \alpha ^2 B^2}{l^2 (\alpha  B+1)^4} \ln \left(\frac{r}{\alpha  l (B+r)+l}\right)~,
\end{equation}
while the function $f(r)$ yields
\begin{multline}
f(r) =-\frac{4}{l^2}+\frac{6 \alpha  B^2 \left(\alpha  \left(\alpha  B^2 (3 \alpha  r-2)+B (\alpha  r (2 \alpha  r-3)-4)-2 r (2 \alpha  r+3)\right)-2\right)}{l^2 r (\alpha  B+1)^3 (\alpha  (B+r)+1)^2}+ \\ \frac{12 \alpha ^2 B^2}{l^2 (\alpha  B+1)^4}\left( (\alpha  B-2) \ln \left(\frac{r}{\alpha  l (B+r)+l}\right)+\alpha  B \ln (\alpha  l)\right) ~,
\end{multline}
The Ricci scalar is divergent at origin and related to the AdS scale at infinity. The Kretschmann scalar behaves as
\begin{equation}R^{\alpha\beta\gamma\delta}R_{\alpha\beta\gamma\delta}(r \to 0) \sim \frac{24 B^6}{l^4 r^6 (\alpha  B+1)^2} + \mathcal{O}\left(\frac{1}{r^4}\right)~,\end{equation}
near the origin, which is also divergent at $r=0$, indicating a physical singularity. It is clear that we cannot invert the Ricci scalar and solve it for $r$, to substitute back to the $f(r)$ function in order to obtain $f(R)$. However, we can use asymptotics to have a feeling of the curvature model at the origin and at infinity. The asymptotic expressions of the Ricci scalar near the origin and at infinity are respectively
\begin{eqnarray}
R(r \to 0) &\sim& -\frac{12 \alpha  B^2}{l^2 r (\alpha  B+1)^3} +\mathcal{O}(\ln(r))~,\\
R(r\to \infty) &\sim& 6\Lambda_{\text{eff}}-\frac{3 B^2}{\alpha ^2 l^2 r^4} + \mathcal{O}(r^{-5})~,
\end{eqnarray}
so the $f(R)$ function near the origin and at large distances yields respectively
\begin{eqnarray}
f(R(r\to 0)) &\sim& R-\frac{12 \alpha^2  B^2 }{l^2 (\alpha  B+1)^3}\ln (R)~,\\
f(R(r\to \infty)) &\sim& R-\frac{4 B \left(6 \Lambda _{\text{eff}}-R\right){}^{3/4}}{3^{3/4}
   \sqrt{\alpha  B l}}~,
\end{eqnarray}

up to a constant of integration. The argument of the $\ln$ tern is not dimensionless, but this expression is an approximation.

In FIG.\ref{f(R)par} we plot $f(R(r))$ as a function of $R(r)$ to see how our $f(R)$ deviates from the GR case $f(R) = R+2l^{-2}$. We can see that for stronger $\alpha$, our theory deviates more from GR.

\begin{figure}
\centering
 \includegraphics[width=0.40\textwidth]{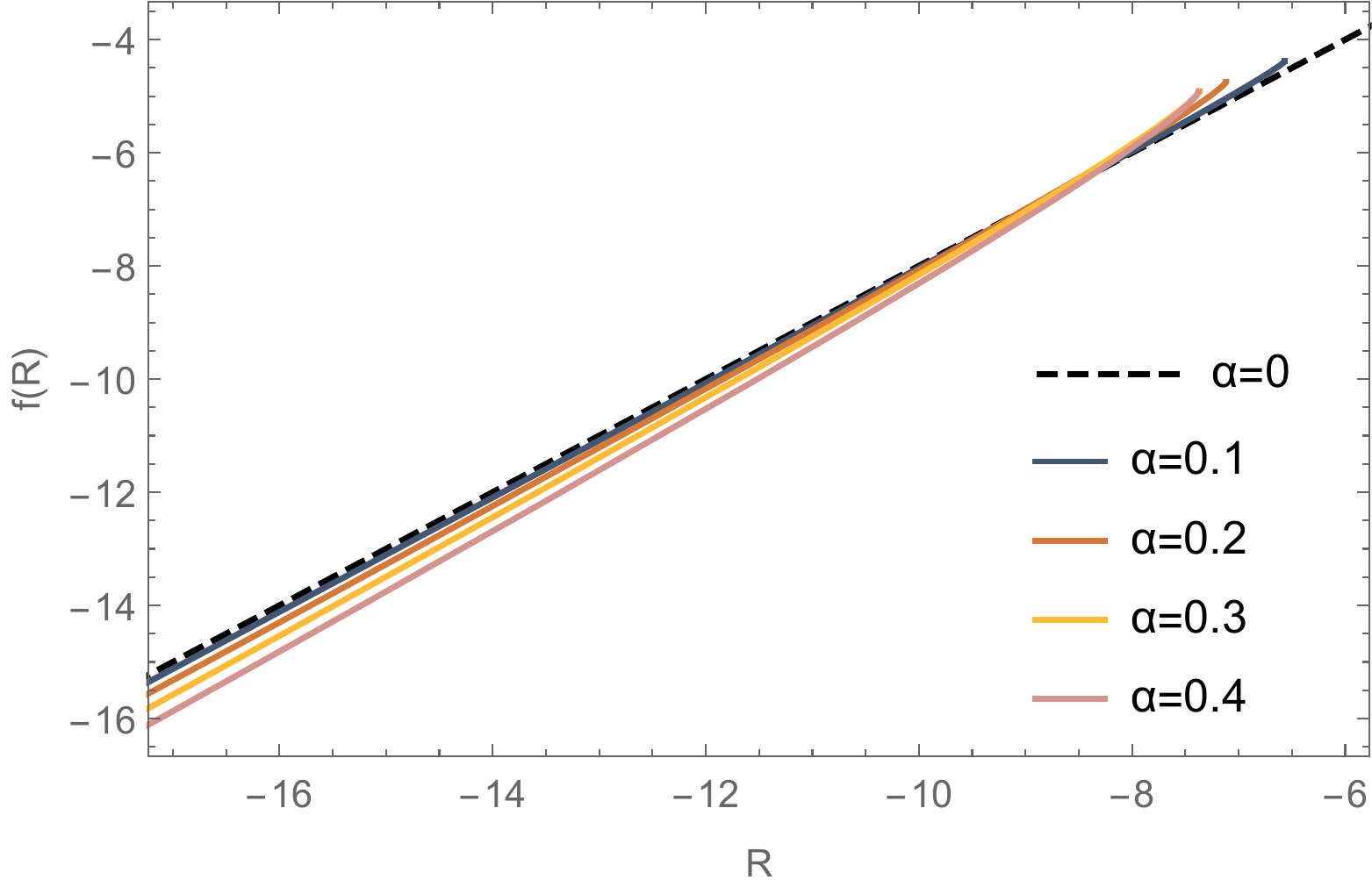}
\caption{$f(R(r))$ as a function of $R(r)$ for different values of $\alpha$, where we have set $B=l=1$.}\label{f(R)par}
\end{figure}

To check if the resultant $f(R,\phi)$ theory is free of ghost and tachyonic instabilities \cite{DeFelice:2010aj,Bertolami:2009cd,Faraoni:2008mf,Faraoni:2006sy,Pogosian:2007sw} we need to confirm if the following relations hold respectively
\begin{eqnarray}
&&f_{R_{\text{total}}} >0 \to f_{R_{\text{total}}} = f_{R_{\text{gravity}}}+ f_{R_{\text{matter}}} = 1+\alpha r - \frac{1}{8}\phi(r)^2 = +1 +\alpha  r -\frac{B}{B+r} = r \left(\alpha +\frac{1}{B+r}\right) >0~,\\
&&f_{RR}(r) >0 \to f_{RR}(r)= \frac{f_{R}'(r)_{\text{total}}}{R'(r)} = \frac{l^2 r^2 (\alpha  (B+r)+1)^3 \left(\alpha  (B+r)^2+B\right)}{12 \alpha  B^2 (B+r)^2}>0~.
\end{eqnarray}

The two above relations hold for $B>0$ and $\alpha>0$ and in this case the resultant $f(R,\phi)$ theory is free of ghosts and avoids the tachyonic instability. The fact that  $f_{R_{\text{total}}} >0$ also ensures that the entropy is positive \cite{Zheng:2018fyn,Camci:2020yre,Akbar:2006mq,Zhu:2020hte} and our solutions may possess a higher entropy than the corresponding GR counterpart \cite{Martinez:1996gn} for a fixed horizon value, meaning that the $f(R)$ black holes may be thermodynamically preferred over the GR one \cite{Martinez:1996gn}.

Expanding the metric function $b(r)$, $f(r)$ and $R(r)$ near $\alpha \to 0$ we find that
\begin{eqnarray}
&&b(r) = \frac{-3 B^2 r-2 B^3+r^3}{l^2 r}+\frac{2 \alpha  \left(3 B^2 r^2+3 B^3 r+B^4\right)}{l^2 r}+\mathcal{O}\left(\alpha ^2\right)~,\\
&&f(r) = -\frac{4}{l^2}-\frac{12 \alpha  B^2}{l^2 r}+\mathcal{O}\left(\alpha ^2\right)~,\\
&&R(r) = -\frac{6}{l^2}-\frac{12 \alpha  B^2}{l^2 r}+\mathcal{O}\left(\alpha ^2\right)~,
\end{eqnarray}

where, as expected, at zeroth order we obtain the GR black hole \cite{Martinez:1996gn} and the curvature functions $f(r), R(r)$ become dynamical due to the gravitational scale $\alpha$.
The trace of the resultant energy-momentum tensor is dynamical
\begin{multline}T_{\mu}^{~\mu} = -\frac{3 \alpha ^2 B^3}{2 r (\alpha
   B+1)^4 (\alpha  l (B+r)+l)^2} \Bigg((\alpha  B+1) (\alpha  (2 \alpha  B^2+B (9 \alpha  r+4)+3 r (2 \alpha  r+3))+2)\\+6 \alpha  r
   (\alpha  (B+r)+1)^2 (\ln (\frac{r}{\alpha  l (B+r)+l})+\ln (\alpha  l))\Bigg)~, \end{multline}

which indicates that the theory is not conformally invariant. This scale is the geometric correction parameter $\alpha$. For vanishing $\alpha$, the trace of the energy momentum tensor vanishes as expected since $\alpha \to 0$ gives the GR case \cite{Martinez:1996gn}. We present some plots of $b(r),R(r),f(r),V(r),T^{~\mu}_{\mu}$ in Fig. \Ref{fig:AdSBH1}, in order to better understand our solution alongside the $\alpha =0$ case which corresponds to GR \cite{Martinez:1996gn}.  We can see that the modified gravity parameter affects the dynamics of the curvature related functions, while the GR black hole \cite{Martinez:1996gn} admits a larger horizon radius in comparison with the modified gravity one. We also plot the horizon radius as a function of $\alpha$. The fact that the larger the deviation from the GR solution \cite{Martinez:1996gn} is, the smaller the horizon radius becomes, is in agreement with the $(3+1)$-dimensional case \cite{Karakasis:2021rpn}.
Next, we will briefly discuss scalar perturbations of the obtained spacetime. For this reason, we consider a massless test scalar field $\phi_{0}$ that satisfies its equation of motion \cite{Tang:2019jkn},
\begin{equation} \Box \phi_{0}=0~. \label{pert_scalar} \end{equation}
Transforming the scalar field as $\phi_{0} = r^{-1/2}\varphi_{0}e^{-i\omega_{0}t}$, the Klein-Gordon equation takes the form of a Schrodinger-like one
\begin{equation} \cfrac{d^2\varphi_{0}}{dr^2_*} + (\omega_{0}^2-V_{\text{eff}})\varphi_{0}=0~, \end{equation}
where we expressed this equation using the tortoise coordinate $r_*=\int dr b(r)^{-1}$. The resulting effective potential is complicated, however its asymptotic expression is
\begin{equation} V_{\text{eff}}(r\to \infty) \sim \frac{3\Lambda^2_{\text{eff}}r^2}{4} +\mathcal{O}\left(\frac{1}{r^2}\right)~,\end{equation}
meaning that there is an AdS boundary at infinity constraining the matter fields, regardless of the modified gravity parameter $\alpha$ and the effect it has on $\Lambda_{\text{eff}}$. We checked that, the inclusion of a mass term for the test scalar field does not change the behavior of the resulting effective potential at large distances. Also, no potential well is formed near the horizon of the black hole for both the massive and massless case, as can be confirmed from FIG. \Ref{Veff}, where we plot the effective potential of the massless case, meaning that the test scalar particles are not trapped near the black hole, so, as a result, the spacetime is stable under massless and massive scalar perturbations.

\begin{figure}
\centering
 \includegraphics[width=.40\textwidth]{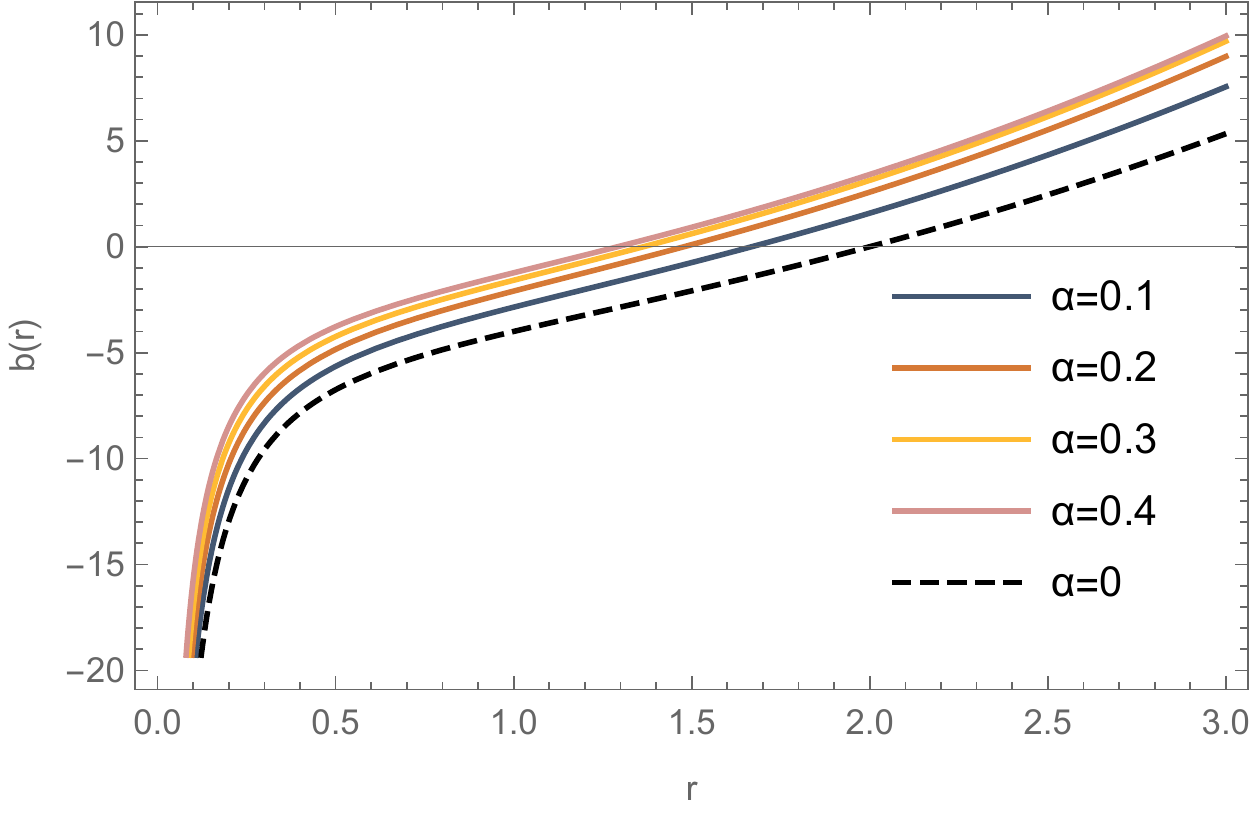}
 \includegraphics[width=.40\textwidth]{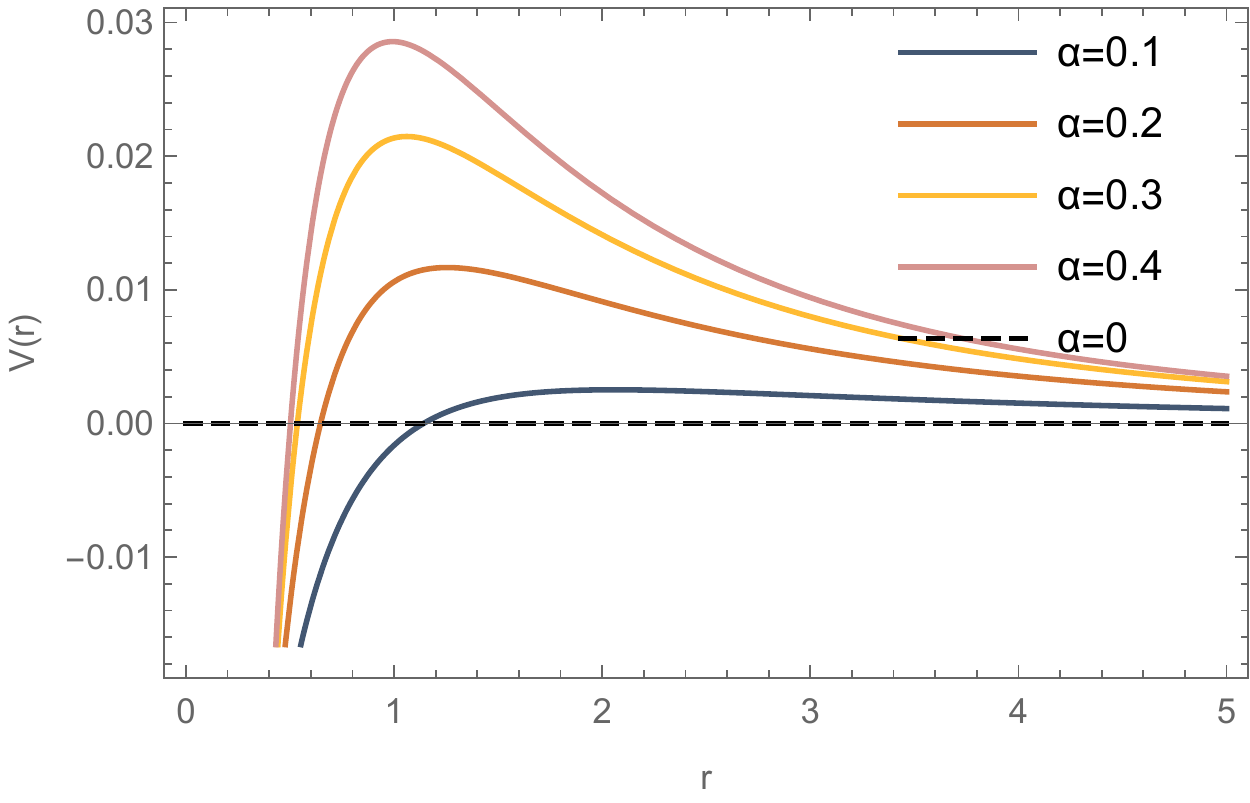}
 \includegraphics[width=.40\textwidth]{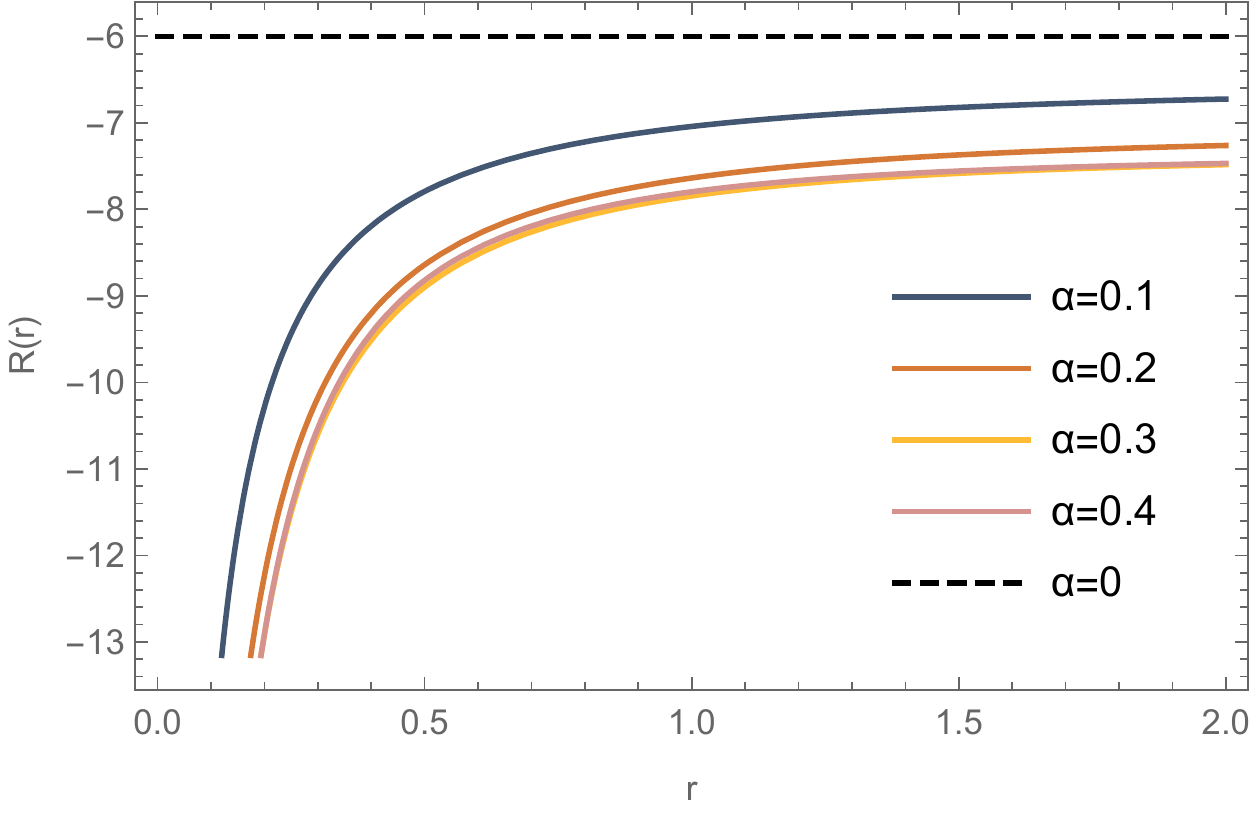}
 \includegraphics[width=.40\textwidth]{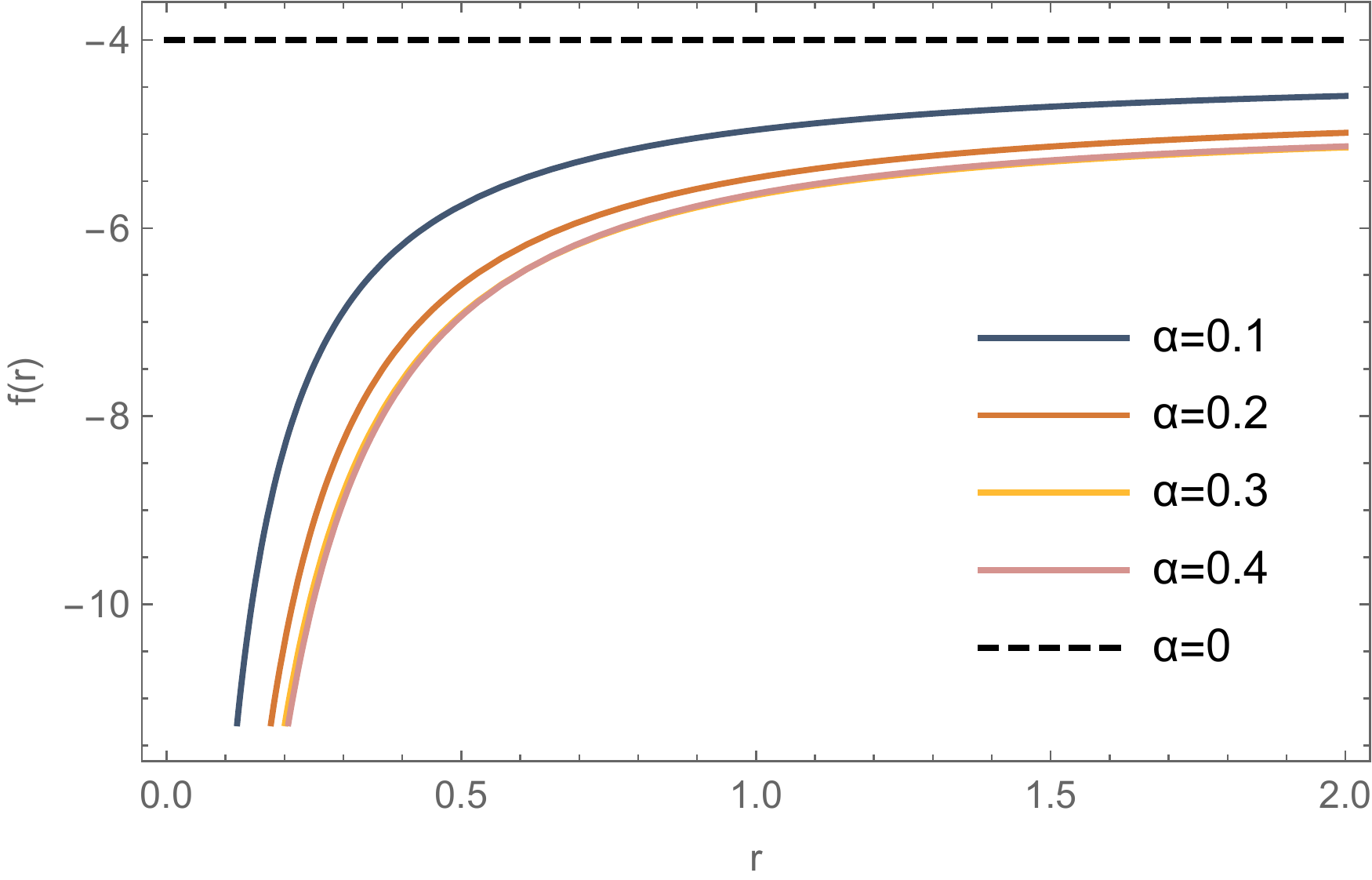}
 \includegraphics[width=.40\textwidth]{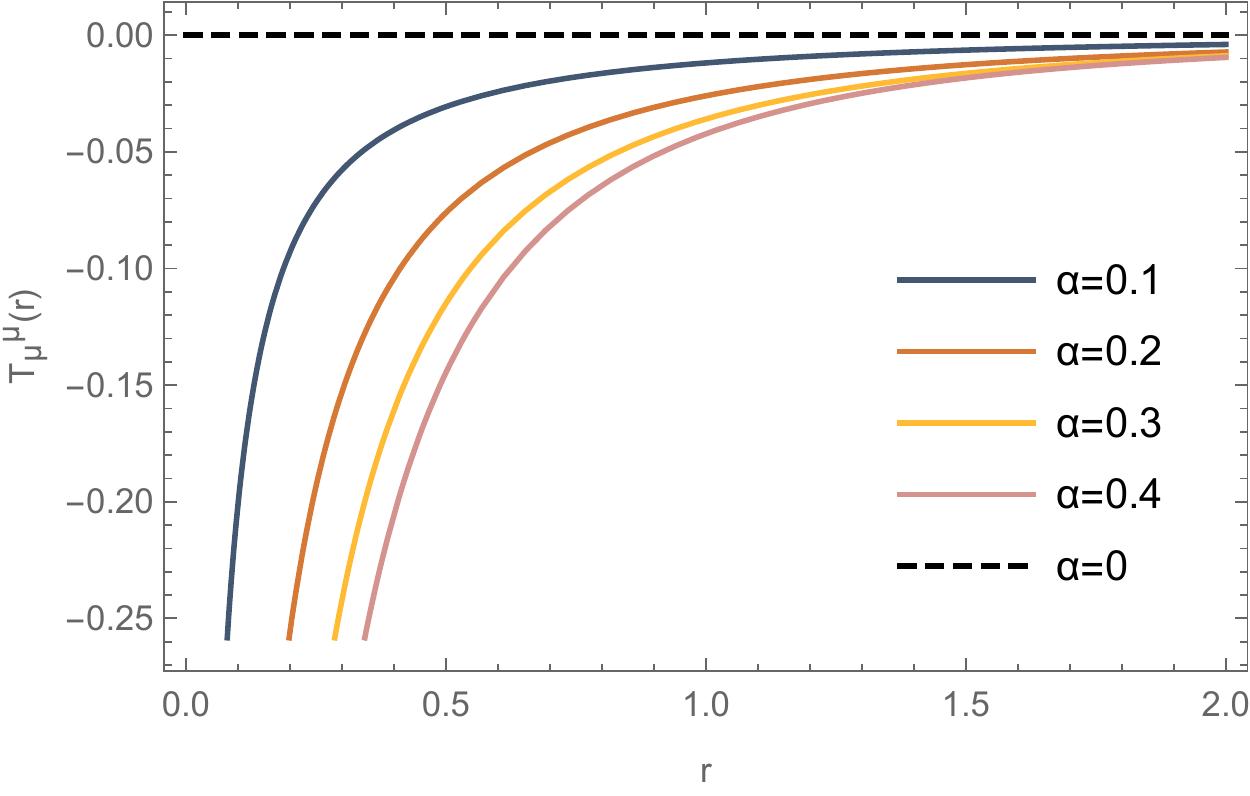}
 \includegraphics[width=0.40\textwidth]{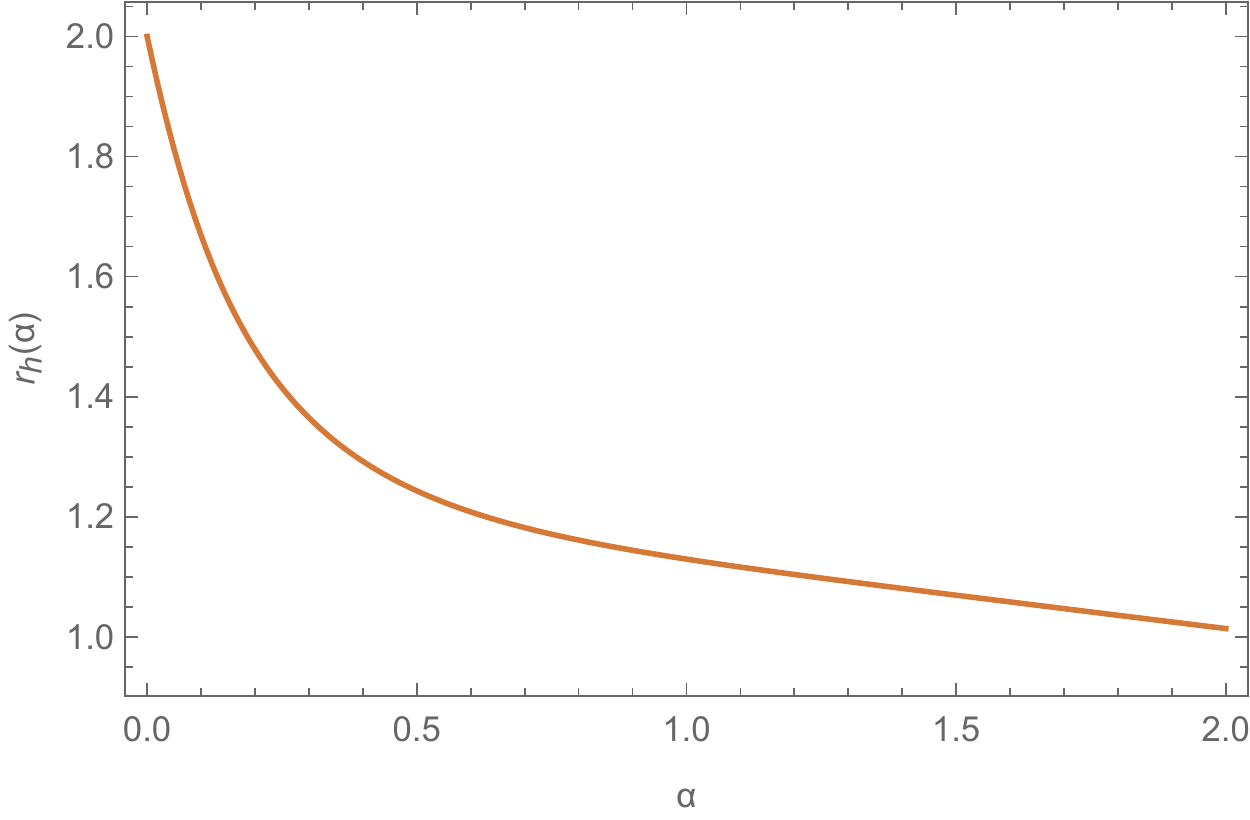}
\caption{The functions $b(r),V(r),R(r),f(r)$ and $T^{~\mu}_{\mu}(r)$ are plotted versus $r$ with different values of $\alpha$, while in the last panel, the radius of the event horizon $r_h$ is plotted as a function of $\alpha$. In all the figures we have set $B=l=1$.} \label{fig:AdSBH1}
\end{figure}

\begin{figure}
\centering
 \includegraphics[width=0.40\textwidth]{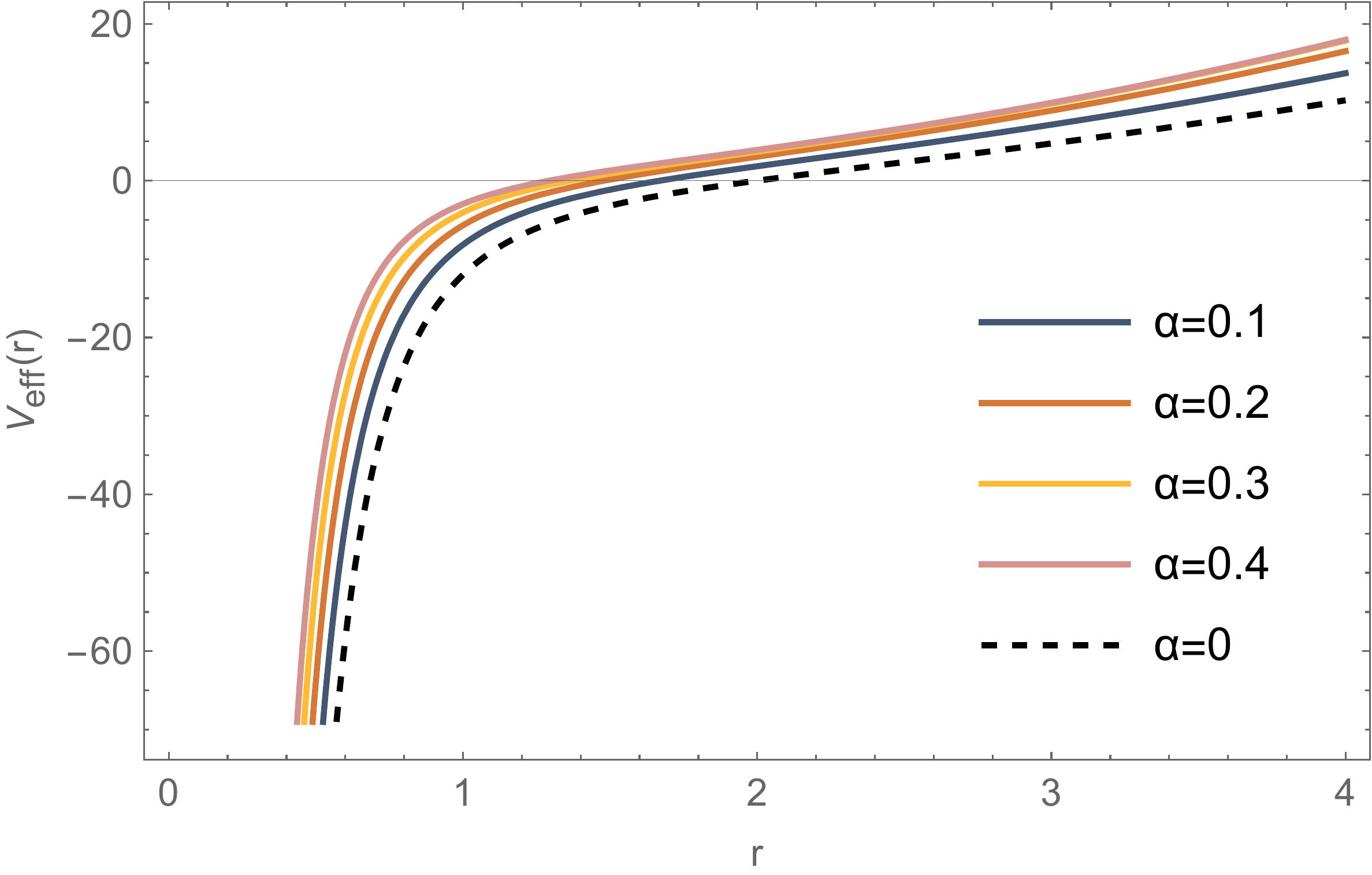}
\caption{The effective potential $V_{\text{eff}}(r)$ for the massless test scalar particles as a function of $r$ for different values of $\alpha$, where we have set $B=l=1$.}\label{Veff}
\end{figure}

\section{Thermodynamics} \label{sect3}

In this section we will study the thermodynamics of the extended black hole solution in $f(R)$ gravity, including Hawking temperature, entropy and the conserved mass.

\subsection{Hawking temperature}

The Hawking temperature can be calculated as
\begin{equation}
    T_H=\frac{b'(r_h)}{4\pi}=\frac{3 B^2 (B+r_h)}{2 \pi  l^2 r_h^2 (\alpha  B+\alpha  r_h+1)}~, \label{frtemp}
\end{equation}
where the relation $b(r_h)=0$ has been used. As expected, it can reduce to the Hawking temperature (\Ref{GRTemp}) in conformal dressed black hole case \cite{Martinez:1996gn} when $\alpha\to 0$.

In FIG. \ref{temperature}, we plot the Hawking temperature $T_H$ as a function of $\alpha$. With the increase of $\alpha$, the Hawking temperature of the black hole first decreases slightly, then grows up to a maximum, finally descends until approaching zero which can be seen from the expression (\ref{frtemp}).

\begin{figure}
\centering
 \includegraphics[width=0.40\textwidth]{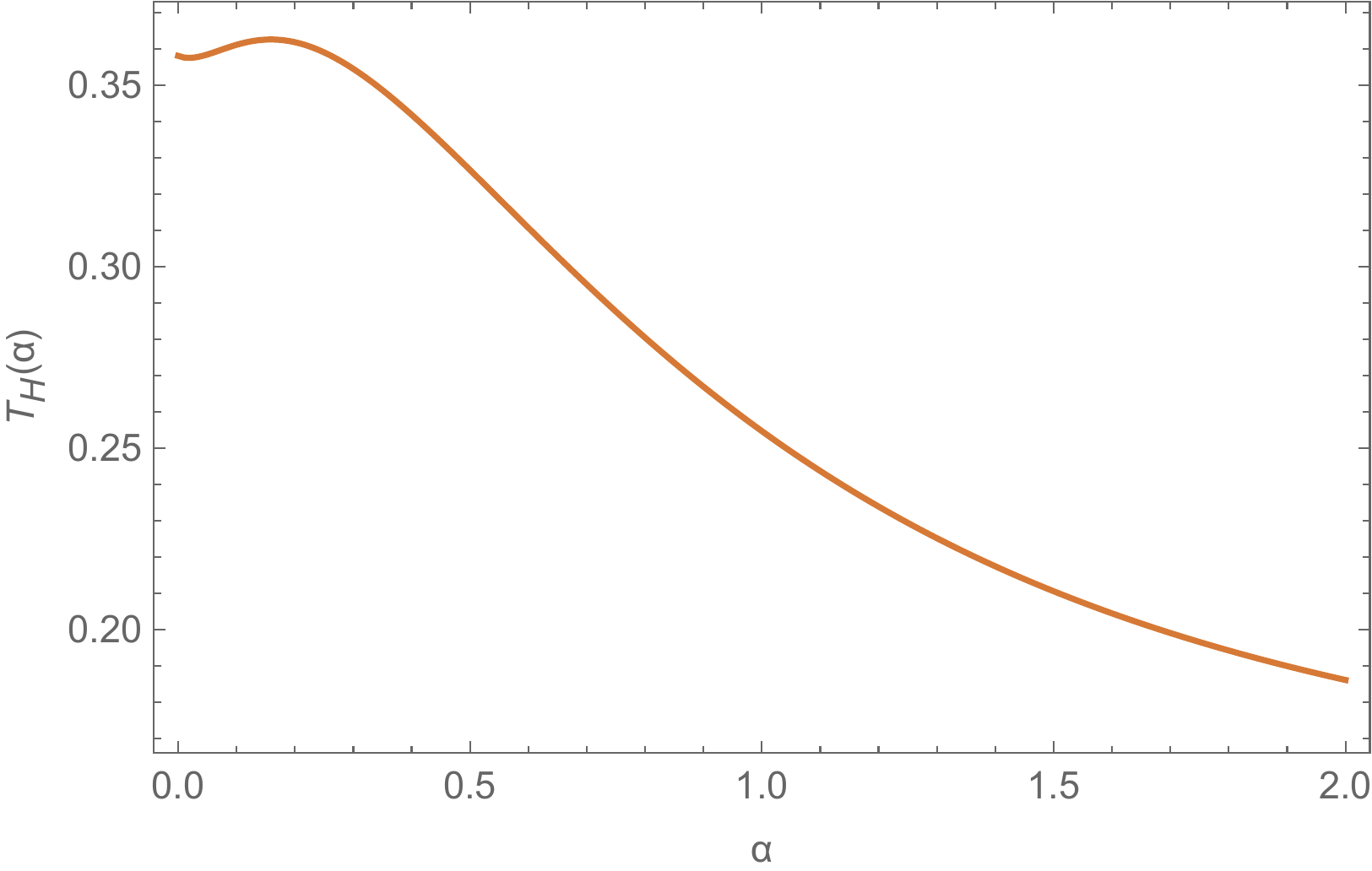}
\caption{The Hawking temperature is plotted as a function of $\alpha$, where we have set $B=l=1$.}\label{temperature}
\end{figure}

\subsection{Entropy}

Using Wald's formula \cite{Wald:1993nt,Iyer:1994ys}, we can calculate the entropy of the black hole in $f(R)$ gravity with a non-minimal coupling as
\begin{equation} S= -\frac{1}{4}\int d\theta \sqrt{r_h^2}\left(\frac{\partial \mathcal{L}}{\partial R_{\alpha\beta\gamma\delta}}\right)\Big|_{r=r_h}\hat{\varepsilon}_{\alpha\beta}\hat{\varepsilon}_{\gamma\delta}~, \end{equation}
where $\hat{\varepsilon}_{\alpha\beta}$ is the binormal to the horizon surface \cite{Dutta:2006vs}, $\mathcal{L}$ is the Lagrangian of the theory, and
\begin{equation} \frac{\partial \mathcal{L}}{\partial R_{\alpha\beta\gamma\delta}}\Big|_{r=r_h} = \frac{1}{2} \left( \frac{f_R(r_h)}{2} - \frac{1}{16}\phi(r_h)^2\right)\left(g^{\alpha \gamma}g^{\beta\delta} - g^{\beta\gamma}g^{\alpha\delta}\right)~.\end{equation}
Finally the formula of the entropy for our theory can be obtained
\begin{equation}S= \pi r_h \left(\frac{f_R(r_h)}{2} - \frac{1}{16}\phi(r_h)^2\right)=\frac{\mathcal{A}}{4}f_{R_{\text{total}}}(r_h)~.\end{equation}

Substituting the explicit expression for $f_{R_{\text{total}}}$, we have
\begin{equation} S= \frac{1}{2} \pi  r_h \left(1+\alpha  r_h-\frac{B}{B+r_h}\right)~,\label{S}\end{equation}
In fact, here $r_h$ is also changing with the choices of $B$, $l$ and $\alpha$.
One might deduce that since $\alpha>0$, the $f(R)$ black holes have higher entropy than the conformal ones \cite{Martinez:1996gn}. However, we have to keep in mind that the conformal case \cite{Martinez:1996gn} has a larger radius for the event horizon as can be seen from the metric function $b(r)$ in Fig.~\Ref{fig:AdSBH1}.
Using the relations (\ref{S}) and $b(r_h)=0$, we can plot the entropy at the event horizon as a function of $\alpha$ in FIG.~\ref{entropy}.

\begin{figure}[H]
\centering
 \includegraphics[width=0.40\textwidth]{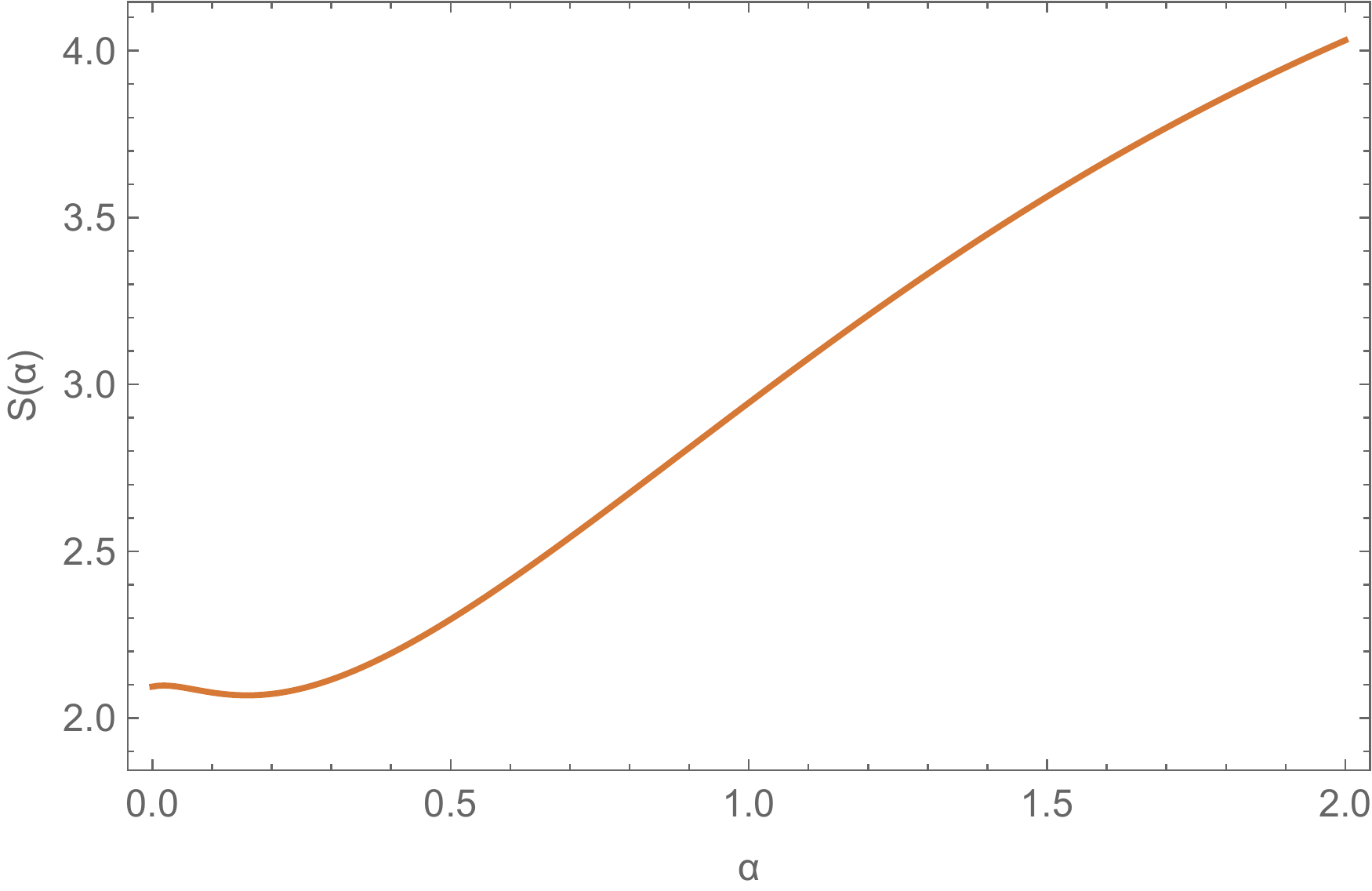}
\caption{ The entropy $S$ at the event horizon is plotted as a function of $\alpha$, where we have set $B=l=1$.
}\label{entropy}
\end{figure}

With the increase of $\alpha$, the entropy first decreases a little bit to a minimum value, then grows up always. Therefore, for most values of $\alpha$, the entropy of the $f(R)$ black hole is higher than the corresponding conformal $(2+1)$-dimensional black hole \cite{Martinez:1996gn}, indicating that our solution is thermodynamically preferred for most cases. It is worth to mention that the conformal case ($\alpha=0$) is a local maximum of the entropy with respect to $\alpha$.

\subsection{Conserved Mass}

For a $D$-dimensional spacetime manifold $\mathcal{M}$, which is topologically the product of a spacelike hypersurface and a real line interval $\Sigma \times \mathcal{I}$, the total quasi-local energy is defined as \cite{Brown:1992br,Brown:1994gs}
\begin{equation}
    E=\int_\mathcal{B} d^{D-2}x \sqrt{\sigma}\varepsilon~,
\end{equation}
where $\mathcal{B}\equiv \partial \Sigma$ is the $(D-2)$-dimensional boundary, $\sigma$ is the determinant of the induced metric $\sigma_{ab}$ on $B$, and $\varepsilon$ is the energy density.

The boundary $\partial\mathcal{M}$ consists of initial and final spacelike hypersurfaces $t'$ and $t''$ respectively, and a timelike hypersurface $\mathcal{T}=\mathcal{B}\times \mathcal{I}$ joining them. The $(D-1)$-metric $\gamma_{ij}$ on $\mathcal{T}$ can be written according to the ADM decomposition as
\begin{equation}
    \gamma_{ij}dx^i dx^j=-N^2 dt^2+\sigma_{ab}\left(dx^a+V^a dt\right)\left(dx^b+V^b dt\right)~.
\end{equation}
The conserved charge associated with a Killing vector field $\xi^i$ is defined as \cite{Brown:1992br,Brown:1994gs}
\begin{equation}
    Q_\xi=\int_\mathcal{B} d^{D-2}x\sqrt{\sigma}\left(\varepsilon u^i+j^i\right)\xi_i~,
\end{equation}
where $u^i$ is the unit normal to spacelike hypersurfaces $t'$ or $t''$, and
$j^i$ is the momentum density.

We first calculate the quasi-local energy inside the spacelike hypersurface $r=r_0=const.$
\begin{equation}
    E=\int_\mathcal{B} d^{D-2}x \sqrt{\sigma}\varepsilon=\int_\mathcal{B} d^{D-2}x \sqrt{\sigma}\left(k-\varepsilon_0\right)~,
\end{equation}
where $k=-\sqrt{b(r_0)}/r_0$ is the trace of the extrinsic curvature and $\epsilon_0$ is the vacuum energy density.

For $r_0\to \infty$, we have the global quasi-local energy
\begin{equation}
    E(r_0)=-\frac{2 \pi  r_0 }{  l}\sqrt{\frac{6 \alpha ^2 B^2 \ln \left(\frac{1}{\alpha  l}\right)}{(\alpha  B+1)^4}+1}-2\pi r_0 \varepsilon_0(r_0)+\mathcal{O}\left(\frac{1}{r_0^2}\right)~.
\end{equation}
To make it finite, the vacuum energy density has to be
\begin{equation}
    \varepsilon_0(r_0)=-\frac{1}{ l}\sqrt{\frac{6 \alpha ^2 B^2 \ln \left(\frac{1}{\alpha  l}\right)}{(\alpha  B+1)^4}+1}~,
\end{equation}
then the global quasi-local energy becomes zero.

The conserved mass can be further calculated as
\begin{eqnarray}
M&=&-\int_\mathcal{B} d^{D-2}x\sqrt{\sigma}\varepsilon u_i \xi^i \notag\\
&=&\lim_{r_0\to\infty}E(r_0)\sqrt{b(r_0)} \notag \\
&=&\lim_{r_0\to\infty}\frac{2 \pi  B^2}{\alpha  l^2 r_0}-\frac{3 \pi  B^2}{2 r_0^2 \alpha ^2  l^2}+\mathcal{O}\left(\frac{1}{r_0^4}\right) \notag \\
&=&0~,
\end{eqnarray}
which, however, turns out to be zero.

The fact that the conserved mass is zero has its root in the $f(R)$ theory. It is known that the conserved mass is related to the constant term in the metric function that survives in the asymptotic expansion at infinity when one is dealing with (A)dS spacetime in $(2+1)$ dimensions. We can split the metric function in two parts. A part that is not completely supported by the gravitational scale $\alpha$ denoted by $b(r)_{GR,\alpha,\phi}$ and a part that is completely supported by $\alpha$, i.e, when we turn off $\alpha$ these terms will vanish, denoted $b(r)_{\alpha,\phi}$. We have $b(r) =b(r)_{\alpha,\phi} + b(r)_{GR,\alpha,\phi}$, where
\begin{eqnarray}
&&b(r)_{GR,\alpha,\phi} = -\frac{3 B^2}{l^2 (\alpha  B+1)^2}-\frac{2 B^3}{l^2 r (\alpha  B+1)} + \frac{r^2}{l^2}~, \label{expr1} \\
&&b(r)_{\alpha,\phi} = \frac{6 \alpha  B^2 r}{l^2 (\alpha  B+1)^3} +r^2\frac{6 \alpha ^2 B^2 }{l^2 (\alpha  B+1)^4}\ln \left(\frac{r}{\alpha  l (B+r)+l}\right)~.\label{expr2}
\end{eqnarray}
It is clear that by setting $\alpha=0$ in $b(r)_{\alpha,\phi}$, the term will vanish, while $b(r)_{GR,\alpha,\phi}$ will yield the conformal black hole solution \cite{Martinez:1996gn}. It can be seen that the $b(r)_{GR,\alpha,\phi}$ part contains a term that is related to the mass of the black hole
\begin{equation} M_{GR,\alpha,\phi} = \frac{3 B^2}{l^2 (\alpha  B+1)^2}~,\end{equation}
while expanding the $b(r)_{\alpha,\phi}$ term at infinity, we find that the constant term will be related to the mass of the black hole reads
\begin{equation}
M_{\alpha,\phi} = - \frac{3 B^2}{l^2
   (\alpha  B+1)^2}~,
\end{equation}
which is the opposite of the mass term the $b(r)_{GR,\alpha,\phi}$ term generates. Hence the term that exists because of the $f(R)$ function in the metric (\Ref{b}), $b(r)_{\alpha,\phi}$ yields a massless black hole, and one can argue that the $f(R)$ theory that satisfies $f_R(r) =s +\alpha r$ yields black holes with no mass. In fact, if one ignores the scalar field and considers only
\begin{equation} S = \int d^3{x}\sqrt{-g} f(R)~, \end{equation}
with our metric ansatz (\ref{metric}) the field equations will naturally yield $f_R(r) = s +\alpha r$, where a logarithmic term that depends on $\alpha$ will cancel the mass the other terms generate yielding massless black holes. For this reason, a more general metric ansatz has to be considered that will yield different profiles for $f_R(r)$, as is indeed recently done in \cite{Nashed:2021jvn}. However, in our case, since we are interested in comparing the $f(R)$ black hole with the GR one \cite{Martinez:1996gn}, we cannot consider a more general metric ansatz, as the metric specifies the form of the scalar field, which further specifies $f_R(r)$ as can be seen in  (\Ref{basic}).

 The parameter $\alpha$ which provides a gravitational correction term to the Ricci scalar term in our $f(R)$ theory, breaks the conformal invariance of the GR case presented in \cite{Martinez:1996gn}. In the case of GR \cite{Martinez:1996gn} the mass of the black hole depends on the scalar charge $B$. In our theory in the metric function (\ref{metric})  both the gravitational parameter and the scalar charge are present and except the mass term there is another term which is proportional to $r^{2}$ which appears in the metric function because of the presence of both the scalar field and the gravitational scale $\alpha$. Considering the expansions of $b(r)_{\alpha,\phi}, b(r)_{GR,\alpha,\phi}$ at infinity in (\ref{expr1}) and (\ref{expr2}) we can say that the massless black hole is a result of the cancellation from the scalar field and the gravitational field contributions. A similar behavior was found in \cite{Martinez:2006an}. Breaking the conformal invariance of the action of the MTZ black hole in the Einstein frame through a particular scalar potential, a massless black hole was found and this was attributed to the cancellation of gravitational and scalar field contributions.

\section{Conclusions} \label{sect4}

In this work, we considered $f(R)$ gravity theory and matter in the form of a self-interacting, non-minimally coupled scalar field. Solving the field equations we found that $f_R(r) =\frac{df(R)}{dR} =1 +\alpha r$ where $\alpha $ is a non-linear correction term of the Ricci scalar $R$, having dimensions  of inverse length. If $\alpha=0 $ we go back to GR recovering the theory of a conformally coupled scalar field to gravity, discussed in \cite{Martinez:1996gn}. The parameter $\alpha $ introduces a gravitational scale that breaks the conformal invariance. Calculating the exact forms of the
derivatives of $f(R)$ function we deduced that $f_{R_{\text{total}}}>0$ and $ f_{RR}>0$ which makes our theory free of ghost and tachyonic instabilities. We also calculated the conserved mass of the black hole and interestingly we found that the black hole is massless due to  the cancellation of gravitational and scalar field contributions to the mass term. We attributed this effect to the breaking of the conformal invariance due to the presence of the gravitational parameter $\alpha$.

We also  studied  the thermodynamics of the extended black hole solution in $f(R)$ gravity, including Hawking temperature and the  entropy.  With the increase of $\alpha$, the Hawking temperature of the black hole first decreases slightly, then grows up to a maximum, finally descends until approaching zero, while the entropy first decreases  to a minimum value, then grows up with the increase of $\alpha$. Besides, the entropy of the black hole is higher than the corresponding conformal $(2+1)$-dimensional black hole \cite{Martinez:1996gn} for most values of $\alpha$, indicating that our solution is thermodynamically preferred for most cases. We also briefly discussed the stability of the obtained spacetime under massive and massless scalar perturbations and deduced that the obtained solution is stable under both types of perturbations.

A possible extension of this work is to perform a detailed thermodynamical analysis  to examine the validity of the first law of thermodynamics, as well as the thermodynamical stability and possible phase transitions of the obtained black hole solution. One can also introduce a linear Maxwell field in the action and study the interplay of the gravitational parameter $\alpha$ and the charge $Q$ on the conformal invariance of the theory and see their effects on the black hole solution. With the addition of electric charge one can also study possible thermodynamical critical behaviors, pointing out how the gravitational scale $\alpha$ affects thermodynamics. Rotating solutions might also be considered. The properties of the resultant conformal field theory could also be studied. The stability of the obtained spacetime may be investigated, as well as the geodesic motion of particles around the black hole solution and how the gravitational parameter $\alpha$ affects the motion.

\end{document}